\documentstyle[11pt]{article}

\begin{document}

\input{epsf}

\def\beq{\begin{equation}}
\def\eeq{\end{equation}}
\def\bea{\begin{eqnarray}}
\def\eea{\end{eqnarray}}
\def\beas{\begin{eqnarray*}}
\def\eeas{\end{eqnarray*}}
\def\ov{\overline}
\def\ot{\otimes}

\newcommand{\hf}{\mbox{$\frac{1}{2}$}}
\def\sig{\sigma}
\def\De{\Delta}
\def\af{\alpha}
\def\be{\beta}
\def\la{\lambda}
\def\ga{\gamma}
\def\ep{\epsilon}
\def\vep{\varepsilon}
\def\half{\frac{1}{2}}
\def\third{\frac{1}{3}}
\def\fth{\frac{1}{4}}
\def\sth{\frac{1}{6}}
\def\tth{\frac{1}{24}}
\def\tde{\frac{3}{2}}

\def\zb{{\bar z}} 
\def\psib{{\bar \psi}} 
\def\etab{{\bar \eta }}
\def\gab{{\bar \ga}}
\def\vev#1{\langle #1 \rangle}
\def\inv#1{{1 \over #1}}

\def\CA{{\cal A}}       \def\CB{{\cal B}}       \def\CC{{\cal C}}
\def\CD{{\cal D}}       \def\CE{{\cal E}}       \def\CF{{\cal F}}
\def\CG{{\cal G}}       \def\CH{{\cal H}}       \def\CI{{\cal J}}
\def\CJ{{\cal J}}       \def\CK{{\cal K}}       \def\CL{{\cal L}}
\def\CM{{\cal M}}       \def\CN{{\cal N}}       \def\CO{{\cal O}}
\def\CP{{\cal P}}       \def\CQ{{\cal Q}}       \def\CR{{\cal R}}
\def\CS{{\cal S}}       \def\CT{{\cal T}}       \def\CU{{\cal U}}
\def\CV{{\cal V}}       \def\CW{{\cal W}}       \def\CX{{\cal X}}
\def\CY{{\cal Y}}       \def\CZ{{\cal Z}}

\newcommand{\np}{Nucl. Phys.}
\newcommand{\pl}{Phys. Lett.}
\newcommand{\prl}{Phys. Rev. Lett.}
\newcommand{\cmp}{Commun. Math. Phys.}
\newcommand{\jmp}{J. Math. Phys.}
\newcommand{\jpamg}{J. Phys. {\bf A}: Math. Gen.}
\newcommand{\lmp}{Lett. Math. Phys.}
\newcommand{\ptp}{Prog. Theor. Phys.}

\newif\ifbbB\bbBfalse                
\bbBtrue                             

\ifbbB   
 \message{If you do not have msbm (blackboard bold) fonts,}
 \message{change the option at the top of the text file.}
 \font\blackboard=msbm10 
 \font\blackboards=msbm7 \font\blackboardss=msbm5
 \newfam\black \textfont\black=\blackboard
 \scriptfont\black=\blackboards \scriptscriptfont\black=\blackboardss
 \def\Bbb#1{{\fam\black\relax#1}}
\else
 \def\Bbb{\bf}
\fi

\def\id{{1\! \! 1 }}
\def\bo{{\Bbb 1}}
\def\bI{{\Bbb I}}
\def\bC{{\Bbb C}} 
\def\bZ{{\Bbb Z}}
\def\CN{{\cal N}}

\title{Multiplicity $A_m$ Models}
\author{{\bf Z. Maassarani}\thanks{Work supported by NSERC 
(Canada) and FCAR (Qu\'ebec).} \\
\\
{\small D\'epartement de Physique, Pav. A-Vachon}\\
{\small Universit\'e Laval,  Ste Foy, Qc,  
G1K 7P4 Canada}\thanks{email address: zmaassar@phy.ulaval.ca} \\}
\date{}
\maketitle

\begin{abstract}
Models generalizing the $su(2)$  XX spin-chain were recently introduced. 
These XXC models also have an underlying $su(2)$ structure. 
Their construction method is shown to generalize to the chains based on 
the fundamental representations of the $A_m$ Lie algebras. 
Integrability of the new  models is 
shown in the context of the quantum inverse scattering method.
Their $R$-matrix is found and shown to yield a representation
of the Hecke algebra.   
The diagonalization of the transfer matrices is carried 
out using the algebraic Bethe Ansatz. I comment on 
eventual generalizations and 
possible links to reaction-diffusion processes.
 
\end{abstract}
\vspace*{2.5cm}
\noindent
\hspace{1cm} May 1998\hfill\\
\hspace*{1cm} LAVAL-PHY-20/98\hfill\\

\vspace*{2.5cm}
\noindent
\hspace*{1cm} PACS numbers: 02.90.+p, 05.50.+q, 75.10.Jm, 82.20.Mj\hfill\\
\hspace*{1cm} Key words: Spin-chain, Integrability, Bethe Ansatz, 
Quantum groups, \hfill\\
\hspace*{1cm} Non-equilibrium Kinetics\hfill\\

\thispagestyle{empty}

\newpage

\setcounter{page}{1}

\section{Introduction}

In the course of generalizing the Hubbard model  new models 
were discovered:  the so-called XXC models \cite{mm,h12,ff,xxc}.
These one-dimensional (spin-chain) integrable models have 
a natural expression in terms of $su(n)$ generators, rather than higher-spin
representations of $su(2)$.  The first hamiltonian of the integrable 
hierarchy is bilinear in terms of $su(n)$ generators. The symmetries 
of the models are generators of $su(n_1)\oplus su(n_2)\oplus u(1)$.
The algebraic Bethe Ansatz diagonalization of the transfer matrices requires 
a nesting similar to the XXZ models based on the fundamental representation
of $su(p)$. Finally, for a  subclass of the 
XXC models the explicit form of all the conserved charges was found
\cite{mm}; their expressions in terms of  $su(n)$ 
structure constants indicate an $su(n)$ interpretation. 
New identities for the structure constants were derived as a by-product.
These should admit further generalizations. 

However the XXC models also share
features associated with the spin-$\frac{1}{2}$ XXZ model and appear 
as a kind of  higher-dimensional representations of the $R$-matrix of 
the spin-$\frac{1}{2}$ model. Indeed, the latter model is a special case of the 
XXC models and  the $R$-matrices of these models  share a common structure, 
with their building blocks satisfying the same algebraic relations.
Moreover the algebraic Bethe Ansatz can also be interpreted as
being nest-less and therefore a simple generalization of the $su(2)$ one. 
Another argument in favor of the $su(2)$ interpretation can be found in
\cite{aars}. One obtains the quadratic  hamiltonian of a particular  XXC model,
with open boundary conditions, as the `infinite-coupling'  restriction
of the Hubbard hamiltonian on a subspace of the complete Hilbert space.
The authors of \cite{aars}  showed that this model possessed
a surprisingly large affine  symmetry based on $su(2)$.
A generalization of the resulting model yielded a subclass of 
open-boundaries XXC hamiltonians; however their symmetries favor an
$su(n)$ interpretation. 

One therefore may try to generalize  the construction
method used in deriving the XXC models, to the XXZ models which are based
on the fundamental representations of the $A_m$ Lie algebras. 
This approach turns out to work and the resulting models are obtained and
studied in this paper.  One  starts with the $A_m$ $R$-matrices. 
Their structure allows a straightforward generalization
retaining their operatorial form and their 
$A_m$ characteristics. The new matrices  are given and shown to satisfy 
the Yang-Baxter equation.
Integrability of the models is then  a simple consequence of the 
quantum inverse scattering framework.
The symmetries  are  obtained and  the transfer matrices 
are diagonalized by algebraic Bethe Ansatz.
I conclude with some remarks and possible physical applications
to reaction-diffusion processes. 

\section{New models}\label{newmod}

The Yang-Baxter equation (YBE) is at the center of 
Quantum Inverse Scattering Method (QISM) used to obtain 
quantum integrable one-dimensional spin-chains and their 
covering two-dimensional classical 
statistical models \cite{qism1,qism2,qism3,jimbo1}.
There are now several methods which can be used to obtain solutions,
$R$- or $L$-matrices, of the YBE. 
An important and quite general method relies on 
the use of affine quantum groups based on
Lie algebras. Rather than directly solve the cubic equations
resulting from the YBE, one solves linear equations where 
$R$ appears as the intertwiner between two possible deformed coproducts
(tensor products). This method  was used  in particular
in \cite{jimbo} and explicit trigonometric solutions were obtained 
for the fundamental representations of the 
classical Lie algebras $A_m$, $B_m$, $C_m$, $D_m$,  and their twisted versions. 

For the untwisted algebra $A_{m-1}$,
the trigonometric $\check{R}$-matrix of the 
fundamental representation is $m^2$-dimensional and can be found 
in \cite{jimbo}:
\bea
\check{R}(y)&=&\sin(\ga) (y \sum_{\af >\be} E^{\be\be}\otimes E^{\af\af}
+ y^{-1} \sum_{\af <\be} E^{\be\be}\otimes E^{\af\af})\nonumber\\ 
&+& \sin(\la+\ga)\sum_{\af}  E^{\af\af}\otimes E^{\af\af}
+\sin(\la)\sum_{\af\not= \be} E^{\be\af}\otimes E^{\af\be}\label{anc}
\eea
where $y=e^{i\la}$, $\la$ is the spectral parameter and $\ga$ the
quantum deformation parameter. 
The  $E^{\af\be}$ are  $m\times m$ matrices  with a one at row $\af$ 
and column $\be$ and zeros otherwise.
The matrix $\check{R}$ satisfies the Yang-Baxter equation
\beq
\check{R}_{12}(\la)  \check{R}_{23}(\la+\mu)\check{R}_{12}(\mu) =
\check{R}_{23}(\mu) \check{R}_{12}(\la+\mu)
\check{R}_{23}(\la)\label{ybec}
\eeq
for any fixed value of $\ga$.
Here and in (\ref{heckeeq}), the notation $\CO_{ij}$ ($i\neq j$)
means that the operator $\CO$ acts non-trivially on the $i^{\rm th}$
and $j^{\rm th}$ spaces, and as the identity on the other spaces. 
The  regularity and unitarity properties also hold:
\beq
\check{R}(0)=\bI\,\sin\ga\;\;,\;\;\; \check{R}(\la) \check{R}(-\la)=\bI\,
\sin(\ga+\la)\sin(\ga-\la) \label{reguni}
\eeq

I now give new solutions to the YBE which are obtained by a multi-state
generalization of expression (\ref{anc}). One first rewrites
the latter matrix as
\beq
\check{R}(\la)=  (y P^{(+)} +y^{-1} P^{(-)})  \sin\ga
+ P^{(2)}\sin(\la+\ga) +  P^{(3)} \sin\la \label{rcm}
\eeq
and looks for representations of the operators $P$ which allow
$\check{R}$ to satisfy the Yang-Baxter equation.
Let $n_i$ be $m$ positive integers such that 
\beq
\sum_{i=1}^m n_i =n \;\;\;\;\;\;{\rm and}
\;\;\;\; 1\leq n_1 \leq ... \leq n_m \leq n -1
\eeq
The inequality restrictions avoid multiple counting of models.
Split the set of $n$ basis states into $m$ disjoint sets $\CA_i$:
\beq
{\rm card}\,(\CA_i) = n_i \;\;,\;\;\; \CA_i \cap  \CA_j= \emptyset 
\;\;\;\;{\rm for}\;\;\; i\not= j 
\eeq
$\CA_i$ should not be confused with the Lie algebra $su(i+1)$. 
Consider the following expression for $P^{(3)}$:
\beq
P^{(3)}=\sum_{1\leq i< j \leq m}\;\sum_{\af_i \in \CA_i}\sum_{\af_j\in \CA_j}
\left(x_{\af_i\af_j} E^{\af_i\af_j}\otimes E^{\af_j\af_i}
+ x_{\af_i\af_j}^{-1} E^{\af_j\af_i}\otimes E^{\af_i\af_j}\right)
\eeq
The twist parameters $x_{\af_i\af_j}$ are arbitrary complex numbers. 
The remaining operators  are given by:
\bea
P^{(1)}&\equiv& (P^{(3)})^2 = P^{(+)} + P^{(-)} =
\sum_{1\leq i< j \leq m}\sum_{\af_i \in \CA_i}\sum_{\af_j\in \CA_j}
\left( E^{\af_i\af_i}\otimes E^{\af_j\af_j}
+ E^{\af_j\af_j}\otimes E^{\af_i\af_i}\right)\label{ppm}\\
P^{(2)}&\equiv&\bI - P^{(1)} = \sum_{i=1}^m \sum_{\af_i\in \CA_i}
\sum_{\be_i\in \CA_i} E^{\af_i\af_i}\otimes E^{\be_i\be_i}\label{p2}
\eea
The operators $P^{(+)}$ and $P^{(-)}$ correspond respectively
to the sums over the first and second summands in (\ref{ppm}).
There is also another way of writing (\ref{rcm}):
\bea
\check{R}(\la)&=& \bI\, \sin(\la+\ga) + P\, \sin\la\\
P&\equiv& P^{(3)} -(e^{-i\ga}\, P^{(+)} +e^{i\ga}\, P^{(-)})
\eea
A straightforward if tedious calculation shows that:
\beq
P^2 = -2\, P\, \cos\ga \;\;,\;\;\; P_{12} P_{23} P_{12} + P_{23} = 
P_{23} P_{12} P_{23} + P_{12}\label{heckeeq}
\eeq
$P$ is therefore a generator of the Hecke algebra.
These relations imply that  the Yang-Baxter equation is satisfied.  
The regularity and unitarity properties (\ref{reguni}) still hold.
I  denote this model by $(n_1, ... , n_m;m,n)$.

There is a simple graphic and mnemonic 
representation of the foregoing operators.
To each state assign a point in the plane. States belonging to the same 
set $\CA_i$ are not linked while those belonging to different sets
are linked. A given link corresponds to a given summand appearing
in the expression of $P^{(3)}$, a tensor product of two step operators,
and also to the summands appearing in $P^{(\pm)}$, a tensor product
of diagonal operators. Similarly the summand  of $P^{(2)}$ 
corresponds to missing
links in the diagram. Links and missing links exhaust all 
possible links which could be drawn between the $n$ states.
This representation is illustrated with two examples in figure 1. 
One can also read this diagram as follows. 
One starts with an $su(m)$  system and replaces  every state
with  an arbitrary number of copies. The copies originating from the 
same state do not `interact'
among each other; they interact with all  other states and their
copies as dictated by the original  diagram.

\vskip 0.5cm
\epsfxsize 16.0 truecm
\epsfbox{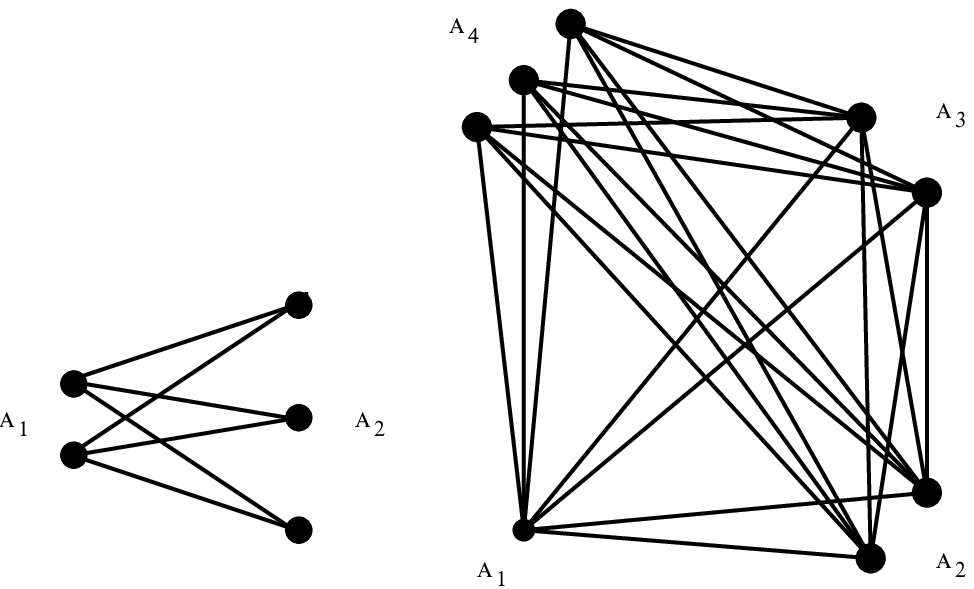}
\vskip 0.5cm
{\small {\bf Fig.1:} The diagram on the left corresponds to
the system $(2,3;2,5)$ and the one on the right to $(1,2,2,3;4,8)$.
Here $A_i$ stands for $\CA_i$.}
\vskip 0.5cm

The choice of which  states go into which set $A_i$ does not
yield inequivalent systems. For instance the two systems,
$(\CA_1=\{ 1\}\,,\; \CA_2=\{ 2,3\})$ and $(\CA_1=\{ 2\}\,,\; \CA_2=\{ 1,3\})$,
are related by a simple (orthogonal) permutation matrix 
whose $N$-fold tensor product with itself yields the  unitary matrix which 
relates the two $N$-sites integrable models \cite{mm}.

The operators  $P^{(1)}$ and $P^{(2)}$ 
form a complete set of projectors on the tensor product space 
$\bC^n\otimes\bC^n$:
\beq
P^{(1)} + P^{(2)}=\bI\;,\;\;  (P^{(1)})^2=P^{(1)}\;,\;\; 
(P^{(2)})^2=P^{(2)}\;,\;\; P^{(1)} P^{(2)}=P^{(2)} P^{(1)}=0
\eeq
One also has $(P^{(3)})^3=P^{(3)}$. However, 
for $m>2$, the operator $P^{(3)}$ does not satisfy the 3-sites relations of  
the `free-fermions' algebra $\CA$ found in \cite{ff}.
This is an important difference between $m=2$ and $m>2$.
In the latter case no `conjugation matrix' exists and it does not 
seem possible to couple two such models. 

For $m=2$ one obtains the XXC models
in their asymmetric guise, as there are factors of $y$ and $y^{-1}$. The transformation to the symmetric models
of \cite{xxc} is given by a simple `gauge' transformation:
\beq 
\check{R}^{GT}(\la)=(A(\la)\otimes \bI)\, \check{R}(\la)\,(\bI\otimes A(-\la))
\label{gtm2}
\eeq
where $A(\la)=\sum_{\af_1} E^{\af_1\af_1} e^{i\la c_1} +\sum_{\af_2}
E^{\af_2\af_2} e^{i\la c_2}$ with $c_2-c_1=1$. 
All the above properties are  preserved by such
a transformation.
For $m>2$ however it is not possible to remove the
$y^{\pm 1}$ factors. 
For $m=n$ and all parameters $x$ equal to one,
one obtains the matrix (\ref{anc}). 

The rational limit of $\check{R}$ is obtained by letting $\la\rightarrow\ga\la$,
dividing by $\sin\ga$ and taking the limit $\ga\rightarrow 0$. These
manipulations conserve all the properties of the $\check{R}$-matrix.
In particular, one obtains
\bea
&\check{R}(\la)=  P^{(1)} +(1+\lambda) P^{(2)} + \lambda  
P^{(3)}\label{ratrcm}&\\
&\check{R}(0)=\bI\;\;,\;\;\; \check{R}(\la) \check{R}(-\la)=\bI\,(1-\la^2)&
\eea

If all parameters $x_{\af_i\af_j}$ are equal to each other, both the trigonometric and rational matrix have the following symmetry
\beq
{[ M\otimes M, \check{R}(\la) ]}= 0\;,\;\;\;\;\;
M=\sum_{k=1}^m M^{(k)}=\sum_{k=1}^m \;\;
\sum_{\af_k,\be_k\in \CA_k} m^{(k)}_{\af_k\be_k} E^{\af_k\be_k}
\eeq
If however  two blocks $M^{(k)}$ and $M^{(k^{'})}$, $k< k^{'}$, 
are diagonal $M$ is 
a symmetry of $R$ with the corresponding parameters $x_{\af_k\af_{k^{'}}}$ 
being  unconstrained. 

The transfer matrix is the generating functional of the infinite set 
of conserved quantities.
Its construction in the framework of the Quantum Inverse Scattering 
Method is well known.
The Lax operator on a chain  at site $i$ with inhomogeneity $\mu_i$
is given by:
\beq
L_{0i} (\la) = R_{0i}(\la-\mu_i) = \CP_{0i} \, \check{R}_{0i}(\la-\mu_i) 
\eeq
where $\CP$ is the permutation operator on $\bC^n\otimes\bC^n$.
The monodromy matrix is a product of Lax operators
\beq
T(\lambda)=M_0\, L_{0N}(\la)...L_{01}(\la)\label{mono}
\eeq
where $N$ is the number of sites on the chain and 0 is the auxiliary
space. 
The transfer matrix is the trace of the monodromy matrix over 
the auxiliary space: $ \tau (\la)= {\rm Tr}_0 \;\left[T(\la)\right]$.
The introduction of $M$ corresponds to integrable periodic $M$-twisted 
boundary conditions. 
A set of local conserved quantities is  given by
\beq
H_{p+1} = \left({d^p \ln\tau (\la)\over d\lambda^p}\right)_{\la=0}
\;\;\; , \;\; p\geq 0 \label{cqs}
\eeq 
The  YBE implies the following intertwining relations for the
elements of the monodromy matrix: 
\beq
\check{R}(\la_1-\la_2) \; T(\la_1)\otimes T(\la_2) =  T(\la_2)\otimes T(\la_1)
\;\check{R}(\la_1-\la_2)\label{rtt}
\eeq
Taking the trace over the auxiliary spaces, and using 
the cyclicity property of the trace, one obtains $[\tau(\la_1),\tau(\la_2) ]=0$.
The hamiltonians $H_p$ therefore mutually commute.

The quadratic hamiltonian calculated from (\ref{cqs}), for $\mu_i=0$
and $M=\bI$, is equal to
\beq
H_2= \sum_j H_{jj+1} = \frac{1}{\sin\ga}
\sum_j \left( P^{(3)}_{jj+1}  + P^{(2)}_{jj+1} \cos \ga
+ (P^{(+)}_{jj+1} - P^{(-)}_{jj+1})\, i\sin\ga \right) \label{h2}
\eeq 
For $|x_{a\beta}|=1$ and $\ga$ purely imaginary the hamiltonian
$\sin\ga \times H_2$ is hermitian.
For $m=2$ this hamiltonian is also hermitian for real values of $\ga$.
Under the periodic boundary conditions the non-hermitian part 
does not contribute. This is easily seen from the  transformation
(\ref{gtm2}) and the fact that the hamiltonian density
$H_{jj+1}$ is equal to the derivative at zero of  $\check{R}(\la)$.
The rational limit yields: $H_2=\sum_j ( P^{(3)}_{jj+1} + P^{(2)}_{jj+1})$,
and, provided $|x_{a\beta}|=1$, the hamiltonians are hermitian. 

The cubic conserved quantity is obtained  from (\ref{cqs}) by a direct
calculation.
One finds $H_3 = -\sum_j {[H_{j-1j},H_{jj+1} ]} - \frac{N}{\sin^2\gamma} \bI$,
where $H_{jj+1}$ is the hamiltonian density of (\ref{h2}) or its rational
version. The $\sin^2\gamma$ is replaced by 1 for the 
rational limit. The commutator
can be easily derived using $E^{\alpha\beta} E^{\gamma\delta}= 
\delta_{\be\ga} E^{\af\delta}$. 
No  general closed form expressions for the higher conserved 
quantities have yet been derived in the literature; however see \cite{mm}
for some specific  cases,  and the references therein for related issues. 

The transfer matrix, and therefore all the conserved quantities,
have the symmetries of the $\check{R}$-matrix. Define the magnetic-field
operators as:  
\beq
H_1^{\af_k\be_k}\equiv \sum_i E_i^{\af_k\be_k}\;\;\;,\;\;\; 
\af_k,\,\be_k\in \CA_k
\eeq
One has the following commutation relations for both the trigonometric and 
rational forms:
\bea
&{[H_1^{\alpha_k\beta_k},\tau(\lambda)]}=0 \;\;\;\;{\rm if\; and\; only \;if}&\\
&\forall j < k\,,\; \forall \ga_j \in \CA_j\;\;
x_{\ga_j\af_k}=x_{\ga_j\be_k} \;\;\;\;{\rm and}\;\;\;\;
\forall j > k\,,\; \forall \ga_j \in \CA_j\;\;
x_{\af_k\ga_j}=x_{\be_k\ga_j}& \nonumber\\
&{\rm and }\;\; m^{(k)}_{\af_k\af_k}=m^{(k)}_{\be_k\be_k}\;,\;\;
m^{(k)}_{\ga_k\af_k}=0 \;\;\forall \ga_k\not=\af_k\;,\;\;
m^{(k)}_{\be_k\ga_k}=0 \;\;\forall\ga_k\not= \be_k &\nonumber
\eea
In particular the diagonal operators $H_1^{\af_k\af_k}$ commute with
the transfer matrix without any constraint on the twist parameters,
but with the above constraints on the matrix $M$.
The {\it rational} transfer matrix, with all $x_{\af_i\af_j}$'s 
equal to one, may have additional symmetries.
For $k\not= k^{'}$ one finds
\bea
&{[H_1^{\alpha_k\beta_{k^{'}}},\tau^{\rm rat.}(\lambda)]}=
{[H_1^{\beta_{k^{'}}\alpha_k},\tau^{\rm rat.}(\lambda)]}=0&\\
&{\rm if\; and\; only \;if}\;\; n_k=n_{k^{'}}=1 \;\; {\rm and}\;\; 
m^{(k)}_{\af_k\af_k}=m^{(k^{'})}_{\be_{k^{'}}\be_{k^{'}}}&\nonumber
\eea

In the following we shall concentrate on the case where $M=\bI$ and 
all the $x$'s
are equal to a single parameter $x$, despite the fact some results
hold for generic parameters. 
The full local symmetry then is $su(n_1)\oplus ...\oplus su(n_m)\oplus u(1)
\oplus ... \oplus u(1)$ where there are $m-1$ $u(1)$'s.

\section{Algebraic Bethe Ansatz}

The diagonalization by algebraic Bethe Ansatz of the foregoing  models 
combines features from the diagonalization of the $(1,...,1;m,m)$
models and the XXC models.  We refer the 
reader to \cite{xxc,qism3,fadd} for some details  
and give here the  new features and results.
The $\check{R}$-matrix here is redefined as the  matrix (\ref{rcm}) divided by 
$\sin(\lambda+\gamma)$. This adds $-  N\cot\ga \,\bI$ to the hamiltonian
(\ref{h2}) and $+\frac{N}{\cos^2 \gamma} \bI$ to $H_3$.
For the rational limit one adds $- N \bI$ and 
$+N \bI$ respectively. 

Let $k_0^{(1)}\in \{ 1,...,m\}$ and  $\ga_{k_0^{(1)}}\in 
\CA_{k_0^{(1)}}$ be given. 
The action of all the elements of the monodromy matrix
on  the pseudo-vacuum $||\gamma_{k_0^{(1)}}\rangle \equiv |\gamma_{k_0^{(1)}}\rangle
\otimes ... \otimes|\gamma_{k_0^{(1)}}\rangle$  is easily derived: 
\bea
&T_{\ga_{k_0^{(1)}}\af_k}\, ||\gamma_{k_0^{(1)}}\rangle\not=0\;\;,\;\;\;
T_{\ga_{k_0^{(1)}}\af_{k_0^{(1)}}}\, ||\gamma_{k_0^{(1)}}\rangle\not=0&\\
&T_{\ga_{k_0^{(1)}}\ga_{k_0^{(1)}}}\, ||\gamma_{k_0^{(1)}}\rangle
=||\gamma_{k_0^{(1)}}\rangle\;\;,\;\;\;
T_{\ga_k\ga_k} \, ||\gamma_{k_0^{(1)}}\rangle = \prod_{i=1}^N\left(
x^{{\rm sign}(k-k_0^{(1)})}\frac{\sin(\la-\mu_i)}{\sin(\la-\mu_i+\ga)}
\right)\; ||\gamma_{k_0^{(1)}}\rangle&
\eea
$\forall k\not= k_0^{(1)}$ and 
$\forall \af_{k_0^{(1)}}\not= \ga_{k_0^{(1)}}$. 
All other elements of $T$ annihilate this vector.

Let $C_{\be_k}\equiv T_{\gamma_{k_0^{(1)}}\be_k}$. 
Operator   $C_{\be_k}$
exactly flips a state $|\ga_{k_0^{(1)}}\rangle$ into a state
$|\be_k\rangle$.  Thus these
operators acting on $||\gamma_{k_0^{(1)}}\rangle$ give a linear combination
of states where exactly one state in 
$||\gamma_{k_0^{(1)}}\rangle$ has been 
changed to $|\be_k\rangle$, at every site.
To show this one uses the following relations: 
\bea
{[H_1^{\af_k\af_k},C_{\beta_{k^{'}}}]}&=&\delta_{\af_k\beta_{k^{'}}}
C_{\beta_{k^{'}}}\;,\;\; \forall k,\, \
\forall k^{'}\not= k_0^{(1)}\,,\;\label{zeroth}\\
{[H_1^{\af_{k_0^{(1)}}\af_{k_0^{(1)}}},C_{\be_k}]}&=&0\;,\;\; 
\forall k\not= k_0^{(1)}\,,\;
\forall \af_{k_0^{(1)}}\not=\ga_{k_0^{(1)}}\label{first}\\
{[H_1^{\ga_{k_0^{(1)}}\ga_{k_0^{(1)}}},C_{\be_k}]}&=&-C_{\be_k}\;,\;\;
\forall k\not= k_0^{(1)}\,. \label{last}
\eea
Relation (\ref{last}) also shows that
$C_{\delta^{(1)}_{k_1}}(\lambda_1)...C_{\delta^{(p_1)}_{k_{p_1}}}
(\lambda_{p_1}) 
\; ||\ga_{k_0^{(1)}}\rangle=0$ for $p_1 > N$. Relation (\ref{first})
shows that this same vector has no $|\af_{k_0^{(1)}}\rangle$ state
in it, and (\ref{first}) and (\ref{last}) yield 
for $\af_{k_0^{(1)}}\not= \ga_{k_0^{(1)}}$:
\bea
&T_{\af_{k_0^{(1)}}\af_{k_0^{(1)}}}&C_{\delta^{(1)}_{k_1}}(\lambda_1)...
C_{\delta^{(p_1)}_{k_{p_1}}}(\lambda_{p_1}) \; ||\ga_{k_0^{(1)}}\rangle=
\nonumber\\
& &\delta_{p_1 N}\prod_{i=1}^N x^{{\rm sign}(k_0^{(1)}-k_i)} \frac{\sin(\la-\mu_i)}{\sin(\la-\mu_i+\ga)}\;
C_{\delta^{(1)}_{k_1}}(\lambda_1)...
C_{\delta^{(p_1)}_{k_{p_1}}}(\lambda_{p_1}) \; ||\ga_{k_0^{(1)}}\rangle
\eea
The operators  $C_{\delta_k}$ are therefore candidates for 
writing the eigenvector Ansatz.
In contrast, the operator $T_{\ga_{k_0^{(1)}}\af_{k_0^{(1)}}}$ acting on
$||\gamma_{k_0^{(1)}}\rangle$ changes the state of only the  {\it first} site 
to $|\af_{k_0^{(1)}}\rangle$. 
This is an unusual feature and these operators cannot be used
to write down an eigenvector Ansatz.

One may therefore take as Bethe Ansatz eigenvector
\beq
|\lambda_1,..., \lambda_{p_1}\rangle \equiv
F^{\delta^{(1)}_{k_1},...,\delta^{(p_1)}_{k_1}}
C_{\delta^{(1)}_{k_1}}(\lambda_1)...C_{\delta^{(p_1)}_{k_{p_1}}}
(\lambda_{p_1}) \;||\ga_{k_0^{(1)}}\rangle\label{eigen}
\eeq 
where the parameters
$\lambda_i$ and the coefficients $F$ are to be determined.
The sums run  over all $k_i$ from 1 to $m$ with $k_i\not= k_0^{(1)}$ and
over $\delta^{(i)}_{k_i}$ in $\CA_{k_i}$.

One then applies the transfer matrix on the state $|\lambda_1,...,
\lambda_p\rangle$ and uses the algebraic relations (\ref{rtt}). 
The foregoing procedure, the nested algebraic Bethe Ansatz, 
is a cumbersome but  straightforward generalization
of the one for the usual $(1,...,1;m,m)$ models, {\it i.e.}
the $su(m)$ XXZ model.  The differences come from the sum on the 
multiple states in each set $\CA_i$, 
as already seen on the  initial eigenvector 
Ansatz (\ref{eigen}). The are $m-1$ levels in the nesting and 
diagonalizing the transfer matrix at one level requires diagonalizing 
a new transfer matrix generated by the repeated use of 
relations (\ref{rtt}). This new transfer matrix corresponds
to a system of the above type but with a reduced number of states
and sites. The nesting stops at the last level at  which 
the new transfer matrix is trivially diagonal.
Technical considerations impose a decreasing or increasing
sequence of $k_0^{(1)}$, one being needed for every level of the
Ansatz. 
For the increasing  sequence $(k_0^{(1)}=1, k_0^{(2)}=2,...,m-1)$,  
$\epsilon=-1$ in the eigenvalue and Bethe Ansatz equations while for the decreasing sequence $\epsilon=+1$. 
The sequence of systems appearing for the increasing and decreasing  
sequences are given by
\bea
(n_1,...,n_m;m,n)\; {\rm for}\; p_0=N \;{\rm sites}
\rightarrow (n_2,...,n_m;m-1,n-n_1) \;{\rm for}\; p_1\;{\rm sites}\rightarrow
\nonumber\\
\cdots\rightarrow (n_{m-1},n_m;2,n-n_1-\cdots -n_{m-2})\; 
{\rm for}\; p_{m-2} \;{\rm sites}\\
(n_1,...,n_m;m,n)\; {\rm for}\; p_0=N \;{\rm sites}
\rightarrow (n_1,...,n_{m-1};m-1,n-n_m)\;{\rm for}\; p_1\;
{\rm sites}\rightarrow\nonumber\\
\cdots\rightarrow (n_1,n_2;2,n-n_3-\cdots -n_m) \;{\rm for}\; p_{m-2} \;{\rm sites}
\eea
The sequence of sites is non-increasing:
\beq
N=p_0 \geq p_1 \geq \cdots \geq p_{m-1} \geq 0 \label{sites}
\eeq
The $p_{m-1}$ appears in the diagonalization
of the last system, $(*,*;2,*)$ in the above series;
this  system is an XXC one.
The above procedure is akin to decimating the $A_{m-1}$ Dynkin diagram 
by going from either of its extremities to the other. 

The eigenvalue at one level is related to the eigenvalue
at the following level as follows:
\bea
\Lambda^{(m-k;p_k)}(\la,\{\la_1^{(k)},...,\la_{p_k}^{(k)}\})=
\delta_{p_{k+1}p_k}\,(n_{q_k}-1)\prod_{i=1}^{p_k}\left(\frac{x^\epsilon
\sin(\la-\la_i^{(k)})}{\sin(\la-\la_i^{(k)}+\ga)}\right)\nonumber\\
+\prod_{i=1}^{p_{k+1}}\left(\frac{x^\epsilon
\sin(\la_i^{(k+1)}-\la+\ga)}{\sin(\la_i^{(k+1)}-\la)}\right)
+\prod_{i=1}^{p_k}\left(\frac{x^{-\epsilon}
\sin(\la-\la_i^{(k)})}{\sin(\la-\la_i^{(k)}+\ga)}\right)\times\nonumber\\
\prod_{j=1}^{p_{k+1}}\left(\frac{x^\epsilon
\sin(\la-\la_j^{(k+1)}+\ga)}{\sin(\la-\la_j^{(k+1)})}\right)\times 
\Lambda^{(m-k-1;p_{k+1})}(\la,\{\la_1^{(k+1)},...,\la_{p_{k+1}}^{(k+1)}\})
\label{eval}\\
(\la_1^{(0)},...,\la_N^{(0)})=(\mu_1,...,\mu_N)\;\;,\;\;\; k=0,...,m-2
\nonumber
\eea 
where the subscript $q_k$ appearing in the $\delta$-term is equal to 
$k+1$ ($m-k$) for $\epsilon=-1$ ($\epsilon=+1$) respectively.
The eigenvalue of the  transfer matrix $\tau(\la)$ is
$\Lambda^{(m;p_0)}(\la,\{\la_1^{(0)},...,\la_N^{(0)}\})$. 
The last eigenvalue 
$\Lambda^{(1;p_{m-1})}(\la,\{\la_1^{(m-1)},...,\la_{p_{m-1}}^{(m-1)}\})$
is independent the spectral parameter and of the inhomogeneities.
It is an eigenvalue of the (constant) unit-shift operator
sending the state on site $i$ to  site $i+1$, 
on a lattice of $p_{m-1}$ sites with $n_m$ ($n_1$) states per site 
for $\epsilon=-1$ ($\epsilon=+1$) respectively.

The parameters $\la_i^{(k)}$ appearing at every level are solutions
of the Bethe Ansatz equations:
\bea
\prod_{l_{k+1}=1}^{p_{k+1}}\left(\frac{x^\epsilon 
\sin(\la_{l_{k+1}}^{(k+1)}-\la_i^{(k)}+\ga)}{\sin(\la_{l_{k+1}}^{(k+1)}-
\la_i^{(k)})}\right) 
\prod_{l_k=1,\,l_k\not=i}^{p_k}\left( 
\frac{\sin(\la_i^{(k)}-\la_{l_k}^{(k)}+\ga)}
{\sin(\la_i^{(k)}-\la_{l_k}^{(k)}-\ga)}\right)\nonumber\\
\times\prod_{l_{k-1}=1}^{p_{k-1}}\left( 
\frac{x^{-\epsilon}\sin(\la_i^{(k)}-\la_{l_{k-1}}^{(k-1)})}{\sin(\la_i^{(k)}-
\la_{l_{k-1}}^{(k-1)}+\ga)}\right)=1\;\;\;,\;\;
i=1,...,p_k\;\;,\;\;\; k=1,...,m-2 \label{baes1}
\eea
and
\bea
\Lambda^{(1,p_{m-1})}\times\;
\prod_{l_{m-1}=1,\, l_{m-1}\not=i}^{p_{m-1}}\;\left( 
\frac{\sin(\la_i^{(m-1)}-\la_{l_{m-1}}^{(m-1)}+\ga)}
{\sin(\la_i^{(m-1)}-\la_{l_{m-1}}^{(m-1)}-\ga)}\right)\nonumber\\
\times\prod_{l_{m-2}=1}^{p_{m-2}}\left( 
\frac{x^{-\epsilon}
\sin(\la_i^{(m-1)}-\la_{l_{m-2}}^{(m-2)})}{\sin(\la_i^{(m-1)}-
\la_{l_{m-2}}^{(m-2)}+\ga)}\right)=1\;\;\;,\;\;
i=1,...,p_{m-1}\label{baes2}
\eea
Finally, the coefficients $F^{\delta^{(1)}_{k_1},...,\delta^{(p_1)}_{k_1}}$
are such that $F$ and is an eigenvector of the transfer matrix
$\tau^{(m-1;p_1)}(\la;\la_1^{(1)},...,\la_{p_1}^{(1)})$. 
Note that, as usually happens in the ABA diagonalization,
these equations imply the vanishing of the residues 
of $\Lambda^{(m-k;p_k)}(\la;\{ \la_i^{(k)}\})$ at the $\la_j^{(k+1)}$'s.
This was expected since the transfer matrix  is non-singular at these values of 
the spectral parameter.

All the possible combinations of $p_i$'s satisfying (\ref{sites})
should be considered. For those with a first vanishing $p_{k^{'}}$, equations
(\ref{eval},\ref{baes1}) truncate accordingly 
with $\Lambda^{(m-k^{'};p_{k^{'}})}\equiv 1$ and all products $\prod_1^0$
set to one; equation (\ref{baes2}) is not used.
The BAE clearly display the $A_{m-1}$ Dynkin diagram structure 
when one uses the shifted parameters $\nu_i^{(k)}=\la_i^{(k)}+k\ga/2$.

At every level $l$ of this  diagonalization one has the 
option of choosing among $n_{k_0^{(l)}}$ possible pseudo-vacuum.
To cover the largest number of subspaces of the Hilbert
space of the chain  one should consider all choices. 
This gives distinct eigenvectors but equal eigenvalues. 
This reflects a large degeneracy of the spectrum and is different
from the simple models $(1,...,1;m,m)$. Another difference 
lies in the appearance of the eigenvalue  $\Lambda^{(1,p_{m-1})}$ at the last 
level. As explained in \cite{xxc} the diagonalization 
of the unit-shift operator is in principle
simple. Formulae for the degeneracies of its eigenvalues have been 
derived by  M. Bauer \cite{michel}.

Are  there  eigenstates not obtained by the foregoing procedure?
From the action of the $C$ operators one infers that 
states lying in the subspaces $\CA_k\otimes...\otimes \CA_k$, 
for every fixed $k$,
are not `reached' by the Ansatz. I now fill a gap
in \cite{xxc} and give the action of the transfer matrix on such states.
One easily derives:
\bea
\tau(\la)\; |\af_k^{(1)},..., \af_k^{(N)}\rangle &=&
\left[ \tau^{(n_k,N)} \label{laste}\right.\\
& & \left. + \bI \sum_{k^{'}\not= k} n_{k^{'}}
\prod_{i=1}^N\left(x^{{\rm sign}
(k^{'}-k)}\frac{\sin(\la-\mu_i)}{\sin(\la-\mu_i+\ga)}
\right)\right] |\af_k^{(1)},..., \af_k^{(N)}\rangle \nonumber
\eea
where $\af_k^{(i)} \in \CA_k$ and  $\tau^{(n_k,N)}$ 
is the unit-shift operator for $n_k$ states and $N$ sites. 
Note that for a given $k$, all the states with the same eigenvalue for
$\tau^{(n_k,N)}$ have the same eigenvalue for $\tau(\la)$.

The rational limit of all the above equations is obtained by 
letting $\la_* \rightarrow \ga\la_*$,
$\mu_i \rightarrow \ga\mu_i$,
dividing the eigenvalues by $(\sin\ga)^N$ and taking the limit 
$\ga\rightarrow 0$. 

For vanishing inhomogeneities, 
the $N^{\rm th}$ power of the eigenvalue $\Lambda^{(m,p_0)}(0,\{ 0\})$
is equal to one because one has the unit-shift operator for $N$ sites
and $n$ states. One can verify this using equations 
(\ref{eval},\ref{baes1},\ref{baes2}).

The eigenvalues of the hamiltonians and higher conserved quantities are 
easily derived  by taking the logarithmic derivatives of the
eigenvalue $\Lambda^{(m;p_0)}(\la,\{\la_1^{(0)},...,\la_N^{(0)}\})$
at vanishing spectral parameter. No closed form expressions of these
derivatives are known. However, for vanishing inhomogeneities $(\mu_i=0)$,
the first $N-1$ logarithmic derivatives of the eigenvalues in (\ref{laste})
are easily found to vanish (in both trigonometric and rational cases). 
The first and  second derivatives, $E_2$ and $E_3$,  of the logarithm 
of the eigenvalue $\Lambda^{(m;p_0)}$ can be easily found:
\beq
E_2= \sum_{i=1}^{p_1} \frac{\sin\gamma}{\sin\la^{(1)}_i \sin(\la^{(1)}_i
+\gamma)} \quad , \quad E_2^{\rm rat.}=
\sum_{i=1}^{p_1} \frac{1}{\la^{(1)}_i (\la^{(1)}_i
+1)}
\eeq
and 
\beq
E_3= 2 \sum_{i=1}^{p_1}\frac{\sin\gamma\,\cos\la^{(1)}_i}{\sin^2 \la^{(1)}_i
\sin(\la^{(1)}_i +\ga)}-(E_2)^2\quad ,\quad 
E^{\rm rat.}_3= 2 \sum_{i=1}^{p_1}\frac{1}{(\la^{(1)}_i)^2
(\la^{(1)}_i +1)}-(E_2^{\rm rat.})^2
\eeq

\section{Conclusion}

I have introduced new classes of solutions of the Yang-Baxter equation
and found their symmetries. 
These models appear as hybrids between $su(n)$ XXZ  and 
$su(m)$ XXZ models but share their main characterizing features
with the latter models. The diagonalization 
of the conserved quantities was then done  
using the algebraic Bethe Ansatz procedure.
 
The issue of completeness of the Bethe Ansatz diagonalization,
be it in its algebraic or coordinate form, is still
an open issue for all but the simplest model, the $su(2)$ XX 
model. Various completeness derivations exists, however
they all include some reasonable but unproven element. 
This should not be construed 
as a hindrance to the study of the thermodynamic limit
in the Thermodynamic Bethe Ansatz (TBA) framework
\cite{takasuzu}. As the number of sites becomes large
the Bethe Ansatz equations are transformed into a system
of non-linear integral equations. To do this  
one assumes that the solutions of the BAE take a particular form  
for which their total  number matches the dimensions of the Hilbert  space. 
This form is  true for most of the solutions and the results 
obtained are in complete agreement with other methods used
to study the thermodynamic limit. This is probably due
to the `fact' that the set  of irregular solutions
has vanishing measure. 
 
Thus a detailed study of the spectrum in  the TBA framework  
is desirable.  
Taking the logarithm of the Bethe Ansatz equations shows that 
the distribution of integers characterizing the solutions will 
get a contribution from the $\Lambda^{(1,p_{m-1})}$ correction
and degeneracies associated with the integers $n_k$.
Whether this influences the central charge and the conformal 
weights remains to be seen. 

Another open issue is the determination of a quantum group
framework. A step in this direction was 
taken in \cite{aars}. It should admit generalizations
and  would shed some light on  whether 
multistate generalization exist for higher representations
of $A_m$ or for other Lie algebras.  
In particular,  the multi-states $A_m$ models should extend
straightforwardly to the Lie superalgebras $su(m|n)$.   

Reaction-diffusion processes in one dimension are described by a time-dependent
probability distribution $P(\{\beta\},t)$ for the configuration $\beta$. 
This distribution obeys a stochastic master equation which can be written
as a Schr\"odinger equation with imaginary time, 
$\partial_t |P(t)\rangle = - H |P(t)\rangle $ and 
$P(\{\beta\},t)$ are the components of the wave function in the basis of 
(species) states of the Hilbert space. 
The hamiltonian $H$ contains the physical transition rates.
For certain processes $H$ was found to belong to integrable hierarchies
described by a Hecke algebra \cite{ar,adhr}. These hierarchies 
enter the class of models studied here.
As the multi-states models are realizations  
of the Hecke algebra,  possible physical applications
may lie in the field of reaction-diffusion
processes of multiple species. 
It would be interesting to pursue such an approach. 

\bigskip\ {\bf Acknowledgement:} I thank D. Arnaudon
for bringing to my attention reference \cite{aars} and for many 
interesting discussions. I also thank M. Bauer for his lightning
derivation of  the complicated  eigenvalue degeneracy formula, 
and the Service de Physique Th\'eorique of Saclay for their hospitality.  
I am grateful for the continued support of  P. Mathieu.

\end{document}